\documentclass[aps,prl,reprint,twocolumn,superscriptaddress]{revtex4-1}
\usepackage[utf8]{inputenc}
\usepackage[english]{babel}
\usepackage{float}
\usepackage{amsfonts,amsmath,amssymb}
\usepackage{bm}
\usepackage{graphicx}
\usepackage{color}
\usepackage{tikz}
\usepackage{upgreek}

\DeclareMathOperator{\Ai}{Ai}

\DeclareMathOperator{\EX}{\mathbb{E}}

\newcommand{\Fish}{\mathcal{I}}
\newcommand{\Draw}{\mathcal{D}}
\newcommand{\Like}{\mathbb{L}}
\DeclareMathOperator{\Prob}{\mathbb{P}}

\begin{document}

\title{Quantum interference test of the equivalence principle on antihydrogen}

\author{P.-P. Cr\'epin} \email[]{pierre-philippe.crepin@lkb.upmc.fr}
\affiliation{Laboratoire Kastler Brossel (LKB), Sorbonne Universit\'e, CNRS, 
ENS-PSL Universit\'e, Coll\`ege de France, Campus Pierre et Marie Curie, 75252, Paris, France}
\author{C. Christen} 
\affiliation{Laboratoire Kastler Brossel (LKB), Sorbonne Universit\'e, CNRS, 
ENS-PSL Universit\'e, Coll\`ege de France, Campus Pierre et Marie Curie, 75252, Paris, France}
\author{R. Gu\'erout} \email[]{romain.guerout@lkb.upmc.fr}
\affiliation{Laboratoire Kastler Brossel (LKB), Sorbonne Universit\'e, CNRS, 
ENS-PSL Universit\'e, Coll\`ege de France, Campus Pierre et Marie Curie, 75252, Paris, France}
\author{V.V. Nesvizhevsky}
\affiliation{Institut Laue-Langevin (ILL), 71 avenue des Martyrs,
 38042, Grenoble, France}
\author{A.Yu. Voronin}
\affiliation{P.N. Lebedev Physical Institute, 53 Leninsky prospect,
117924, Moscow, Russia}
\affiliation{Russian Quantum Center, 100 A, Novaya street, Skolkovo, 
143025, Moscow, Russia}
\author{S. Reynaud} \email[]{serge.reynaud@lkb.upmc.fr}
\affiliation{Laboratoire Kastler Brossel (LKB), Sorbonne Universit\'e, CNRS, 
ENS-PSL Universit\'e, Coll\`ege de France, Campus Pierre et Marie Curie, 75252, Paris, France}
\date{\today}

\begin{abstract}
We propose to use quantum interferences to improve the accuracy of the measurement of the free fall acceleration $\overline{g} $ of antihydrogen in the GBAR experiment. This method uses most antiatoms prepared in the experiment and it is simple in its principle as interferences between gravitational quantum states are readout without transitions between them. We use a maximum likelihood method for estimating the value of $\overline{g}$ and assess the accuracy of this estimation by a Monte-Carlo simulation. We find that the accuracy is improved by approximately three orders of magnitude with respect to the classical timing technique planned for the current design of the experiment. 
\end{abstract}

\maketitle

\paragraph{Introduction - }

Gravitational properties of antimatter raise an important question in the context of the matter-antimatter asymmetry problem
\cite{Bondi1957,Scherk1979,Nieto1991,Chardin2018}. Experimental knowledge on this question is much less precise than for gravitational properties of ordinary matter \cite{Adelberger1991,Darling1992,Huber2000}. For example, the aim of measuring the free fall acceleration  $\overline{g} $ of antihydrogen ($\overline{H} $) in Earth's gravitational field has been approached only recently \cite{TheALPHACollaboration2013} with the sign of $\overline{g} $ not even known yet. Several collaborations are working with antihydrogen atoms produced at CERN to improve the accuracy of $\overline{g}$-measurement in dedicated experiments \cite{Kellerbauer2008,Indelicato2014,Bertsche2017}.

The GBAR collaboration is installing an experiment at CERN, using the techniques of ultracold atom physics to cool down antihydrogen atoms to microKelvin temperatures \cite{Walz2004}. This makes feasible the aim of measuring $\overline{g}$ with an accuracy of the order of 1\% by timing the classical free fall of antiatoms on a well-defined free fall height \cite{Perez2015,Mansoulie2019}.
In this paper, we propose to improve the accuracy of this measurement with the same cloud of ultracold antihydrogen atoms by using the idea of quantum techniques drawn from experiments performed by inducing transitions between gravitationally bound quantum states (GQS) of ultracold neutrons  \cite{Nesvizhevsky2002nature,Nesvizhevsky2003,Nesvizhevsky2005,Pignol2014}. 

Ultracold neutrons bounce above a matter surface due to the repulsive Fermi interaction \cite{Nesvizhevsky2015}. 
For atoms, quantum bounces may be produced by the rapidly varying attractive van der Waals/Casimir-Polder interaction
above the surface 
\cite{Lennard-Jones1936III,Lennard-Jones1936IV,Berry1972,Yu1993,Berkhout1993,Carraro1998,%
Shimizu2001,Druzhinina2003,Pasquini2004,Friedrich2004,Oberst2005,Pasquini2006,Zhao2008}. 
The mechanism is expected to work with antihydrogen atoms, thus preventing their annihilation at the matter surface \cite{Voronin2005jpb,Froelich2012,Dufour2013,Dufour2015}.

Atoms with a low vertical velocity above the surface are trapped by the combined action of quantum reflection and gravity \cite{Jurisch2006,Madronero2007}. They can thus stay in quantum levitation states for long times which can exceed one second over an helium surface \cite{CrepinEPL2017}. 
With the quantum interference technique studied in this paper, which is inspired by studies of neutron whispering gallery modes \cite{Nesvizhevsky2009}, the transition frequencies between these states are not perturbed by any mechanism inducing transitions. They are well known in the case of perfect quantum reflection, and only submitted to shifts due to the Casimir-Polder interaction which have been precisely calculated \cite{Crepin2017}. It follows that the accuracy of the $\overline{g}$-measurement can be improved by using quantum interference  techniques on these quantum levitation states.

\paragraph{Outlook - }
In this article we propose a new method that consists of measuring the coordinates in space and time of the annihilation of antihydrogen atoms on a detector, thus producing an interference pattern. Similar methods were used in experiments on neutron whispering gallery \cite{Nesvizhevsky2009} and we present here a detailed study of the method applied to antihydrogen atoms. In contrast to previous ideas \cite{Dufour2014shaper}, there is no need for a velocity selection, which allows for a large gain in accuracy while effectively using most antihydrogen atoms.

The new method assumes simultaneous measurement of many gravitational quantum states, thus enormously increasing statistics compared to previous proposals \cite{Voronin2014wag,Nesvizhevsky2019} which considered one or a few quantum states. A practical implementation of this method is also simple, since it does not need precision optics and mechanics and does not need the selection of a single quantum state.

\begin{figure}[bh]
        \center
        \includegraphics[width=1\linewidth]{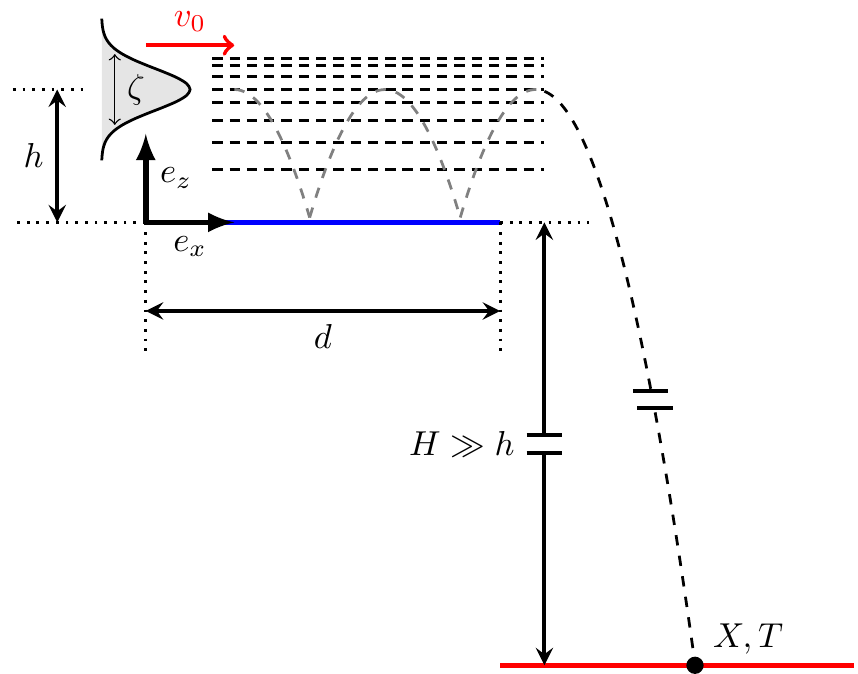}       
        \caption{Schematic representation of the experimental setup. The mirror, of length $d$, is shown as the blue horizontal line and the detector, a distance $H$ below, as the red horizontal line. $X$ and $T$ are the positions in space and time of the detection events ($e_x$-axis horizontal, $e_z$-axis vertical). The wave packet has initially a mean height $h$ above the mirror, a dispersion $\zeta$ and an horizontal velocity $v_0$. The parabolas represent a classical motion with rebounds above the mirror while the horizontal dashed lines represent the paths through different quantum states which interfere in the detection pattern. [Colors on line]}
        \label{Schema}
\end{figure}

We give below a precise description of the new quantum interference technique which should  lead to a largely improved accuracy for the $\overline{g}$-measurement. Starting from the ultracold antihydrogen $\overline{H} ^+$ ions prepared in the GBAR experiment, the method appears as a sequence of steps schematized in figure \ref{Schema}.

First, the $\overline{H} ^+$ wave packet is prepared as the ground state of an ion trap  \cite{Indelicato2014,Perez2015}, submitted to a kick giving it a mean horizontal velocity $v_0$ and irradiated by a laser photodetachment pulse which releases freely falling  $\overline{H}$ atoms. The kick can be produced for example by electrostatic means or by the photodetachment process. 

Atoms then bounce above a surface, treated here as a perfect quantum reflector, and the quantum paths corresponding to different GQS interfere. The interference pattern thus produced is detected after a macroscopic free fall down to an horizontal detection plate. The analysis of the distribution in space and time gives access to the estimation of $\overline{g}$. 

This estimation is in principle sensitive to the initial distribution of the atoms after the kick, and this distribution will have to be determined from the envelop of the detection pattern, as discussed below. In the following, we choose a simple model for this distribution, which is sufficient for the purpose of the present paper.

In figure 1 as well as in the text, lowercase letters represent the quantities relative to first stages of preparation and interference above the mirror, while uppercase letters represent quantities associated with free fall and detection stages. In the following, $\overline{g}$ is simply written $g$ and the standard value $g\simeq9.81\text{m.s}^{-2}$ is used for numerical applications and plots.

\paragraph{Interference of gravitational quantum states - }
Atoms of mass $m$ are released at height $h$ above the perfectly reflecting mirror, 
in a gaussian wave packet factorized along $x$ and $z$ axis $\Psi_0(x,z)=\phi_0(x)\psi_0(z)$
which minimizes Heisenberg uncertainty relation~:
\begin{align}
\label{psi0}
&\psi_0(z)=\left(\frac{1}{2\pi\zeta^2} \right)^{1/4} 
\exp\left(-\frac{(z-h)^2}{4\zeta^2}\right)  ~, \\
&\phi_0(x)=\left(\frac{1}{2\pi\zeta^2} \right)^{1/4} 
\exp\left(-\frac{x^2}{4\zeta^2}+i\frac{mv_0}{\hbar}x \right) ~, \notag
\end{align}
where $\zeta$ is the dispersion of positions, identical along the 2 axis, and $v_0$ the velocity kick. 

Here, we consider the simple model where this distribution is not modified by the kicking mechanism.
As already stated, this point will have to be verified in forthcoming experiments, and this can be done 
by analyzing the envelop of the detection pattern. 
The purpose of the present paper is to evaluate the accuracy which can be obtained
once the initial distribution is known and our simple model is sufficient for this purpose.

The wave packet obeying the Schrödinger equation remains factorized 
 $\Psi_t(x,z)=\phi_t(x)\psi_t(z)$ as long as the atoms
remain in the quantum levitation states above the reflecting surface. 
This condition of separability of the Hamiltonian in $x$ and $z$ coordinates imposes constraints 
on the quality of the mirror surface roughness 
and material homogeneity, and the latter can be met. 

The horizontal evolution leads to a mere spreading of the wave packet. 
The vertical evolution can be decomposed on the orthogonal basis of 
Airy functions ~:
\begin{equation}
\psi_t^n(z) = \Theta(z) \frac{\Ai(z/\ell_g - \lambda_n)}{\sqrt{\ell_g} 
\Ai'(-\lambda_n)} \exp\left(-i\lambda_n t/t_g\right) ~,
\end{equation}
where $\Theta$ is the Heaviside step function describing perfect reflection at the surface, Ai is the first Airy function and $(-\lambda_n)$ its n$^{th}$ zero; $t_g$, $\ell_g$ and $p_g$ are the typical time, length and momentum scales determined by $\hbar$, $m$ and $g$~:
\begin{align}
\label{tg}
t_g&\equiv \left(\frac{2\hbar}{mg^2}\right)^{1/3}\simeq 1.09 \text{ ms}  ~,\\
\label{pg}
p_g&\equiv \frac\hbar{\ell_g} \equiv \left(2\hbar m^2g\right)^{1/3}\simeq1.79\times10^{-29}\,\text{kg.m.s}^{-1} ~.\notag
\end{align}

The wave function $\psi_t(z)$ can be read $\sum_n c_n \psi_t^n(z)$ with
the $c_n$'s obtained by using the orthogonality properties of the $\psi_t^n$-basis. 
When the vertical dispersion of $\psi_t$ is sufficiently small as compared with $h$, they have an analytical expression \cite{Dufour2015}~:
\begin{equation}
\label{cn}
\begin{split}
c_n \simeq {\frac{\zeta^{1/2}}{\ell_g^{1/2}}}\frac{ \left(8\pi\right)^{1/4}}{\Ai'(-\lambda_n)} \Ai\left(\frac{h}{\ell_g}-\lambda_n+\frac{\zeta^4}{\ell_g^4}\right) \times \\
 \exp\left(\frac{\zeta^2}{\ell_g^2} \left(\frac{h}{\ell_g}-\lambda_n+\frac{2}{3}\frac{\zeta^4}{\ell_g^4} \right)\right)~.
 \end{split}
\end{equation}

Parameters in the initial wave packet $\Psi_0$, namely the mean height $h$, standard deviation $\zeta$ and velocity kick $v_0$, have to be chosen so as to optimize the measurement. By tuning the initial Gaussian wave packet up to reasonable values, one can indeed achieve different superposition of states. 
For example, choosing parameters so that the number of low-lying GQS involved remains small leads to a simple interference pattern  with a good contrast on the detector screen. On the contrary, a mixture of a large number of higher-lying GQS states leads to a smaller contrast, because the different states present minima and maxima at different positions. 
We restrict the discussion on the range of parameters that can be experimentally achieved, as known from the existing analysis of the GBAR experiment \cite{Perez2015,Mansoulie2019}. 

We choose for all calculations and plots a length of the reflecting surface d = 5 cm and a height of free fall H = 30 cm,
an initial height above the surface $h=10\ \upmu\text{m}$  and a position dispersion $\zeta = 0.5\ \upmu\text{m}$. 
The last value corresponds to atoms released from an harmonic trap with an oscillation frequency $\omega/(2\pi)= 20$~kHz, that is also a velocity dispersion $\simeq 6.3$~cm.s$^{-1}$ or a zero-point energy $\simeq 42$~peV ($\simeq 0.48  \upmu$K in equivalent temperature units). 

These parameters lead to a large number of interfering GQS which produces patterns 
with high resolution and allows for a good sensitivity to the value of $g$, as discussed below. The final results do not depend in a critical manner on these choices, though the detailed values are in principle affected.  

The high-lying QGS are truncated by putting an absorber at some height above the 
quantum reflecting mirror \cite{Dufour2014shaper}. 
We account for 100 GQS, which  corresponds to an absorbing mirror placed
at $\sim 360~\mu$m above the mirror and leads to $\sim$20\% $\overline{H}$-atoms lost  in the absorber. 

The interaction of wave with a rough surface was studied for neutrons \cite{Voronin2006,Meyerovich2006, Adhikari2007,Escobar2014}, and the physics is the same for the antihydrogen case. The presence of the absorber may induce a couple of extra quantum states (above the absorber height) 
which do not affect significantly the interference pattern which involves many states.

Finally the kick velocity $v_0$ is chosen with the prime criterion that most prepared antiatoms are used in the measurement. This requires that $v_0$ exceeds the standard deviation of the initial velocity distribution by a large enough factor. Another constraint is that most atoms survive their flight above the reflecting surface in a realistic situation where quantum reflection is not perfect. 

Estimations in \cite{CrepinEPL2017} have shown that the best surface for that purpose is liquid helium  
at sufficiently low temperature to suppress the effect of
residual vapors, on which antihydrogen in GQS can bounce for times exceeding one second, 
that is also for a number of bounces exceeding a few hundreds. 
Additionally, a too low velocity results in the final pattern to detections concentrated around the same spot 
on the plate, making an accurate estimation more difficult. 

Several values are chosen below for the kick velocity which satisfy all of the constraints discussed above. Figures are plotted for $v_0 = 0.8\,\text{m.s}^{-1}$ for the sake of a good visibility, while Monte Carlo simulations at the end of the paper
are done with an optimized $v_0 = 0.25\,\text{m.s}^{-1}$.

\paragraph{Readout of the interference pattern - }

The interference pattern produced in the interference zone above the reflecting surface is read out after a free fall 
from the positions in space and time of annihilation events of antihydrogen at the detector. 
Since the height of the free fall is much larger than the position dispersion of the wave packet above the mirror, 
the free fall can be considered as classical (more details below). 
The free fall thus acts in a similar way as a diffraction process, with the \textit{space and time positions} of the annihilation
event on the detector reading the \textit{interaction time and momentum} of the atom leaving the interference zone. 

The description of quantum evolution is thus performed by two different methods in the interference zone and in the free fall zone, with the wave function matched at the virtual surface separating the two zones.  This matching corresponds to the continuity of the wave function (and its derivative) at the surface $x=d$ with $\Psi(d^-,z)=\Psi(d^+,z)$, so that we can write $\Psi(d,z)$ without creating confusion there. 

This treatment amounts to neglect diffraction at the endside of the mirror and  horizontal quantum reflection induced by the change of the potential landscape at $x=d$. The same approximation used in the theoretical description of neutron whispering gallery modes leads to a satisfactory agreement with experiments \cite{Nesvizhevsky2019,Nesvizhevsky2009}. 

This implies that the relevant quantity after the interference zone is the squared wave function in the momentum representation. 
We denote $\widetilde{\psi_t}(p_z)$ the Fourier transform of $\psi_t(z)$ and calculate the relevant signal as probability in momentum space of the wave function~:
\begin{align}
& \Pi_t(p_z) = |\widetilde{\psi_t}(p_z)|^2 = \sum_{n,m} c_n c_m^* \pi_{n,m}(p_z) e^{i \omega_{nm} t} ~, \notag\\
&	\pi_{n,m}(p_z) \equiv \widetilde{\psi_n}(p_z)\left(\widetilde{\psi_m}(p_z) \right)^* ~.
\label{momentumpro}
\end{align}

\begin{figure}[ht]
\centering
\includegraphics[width=1\linewidth]{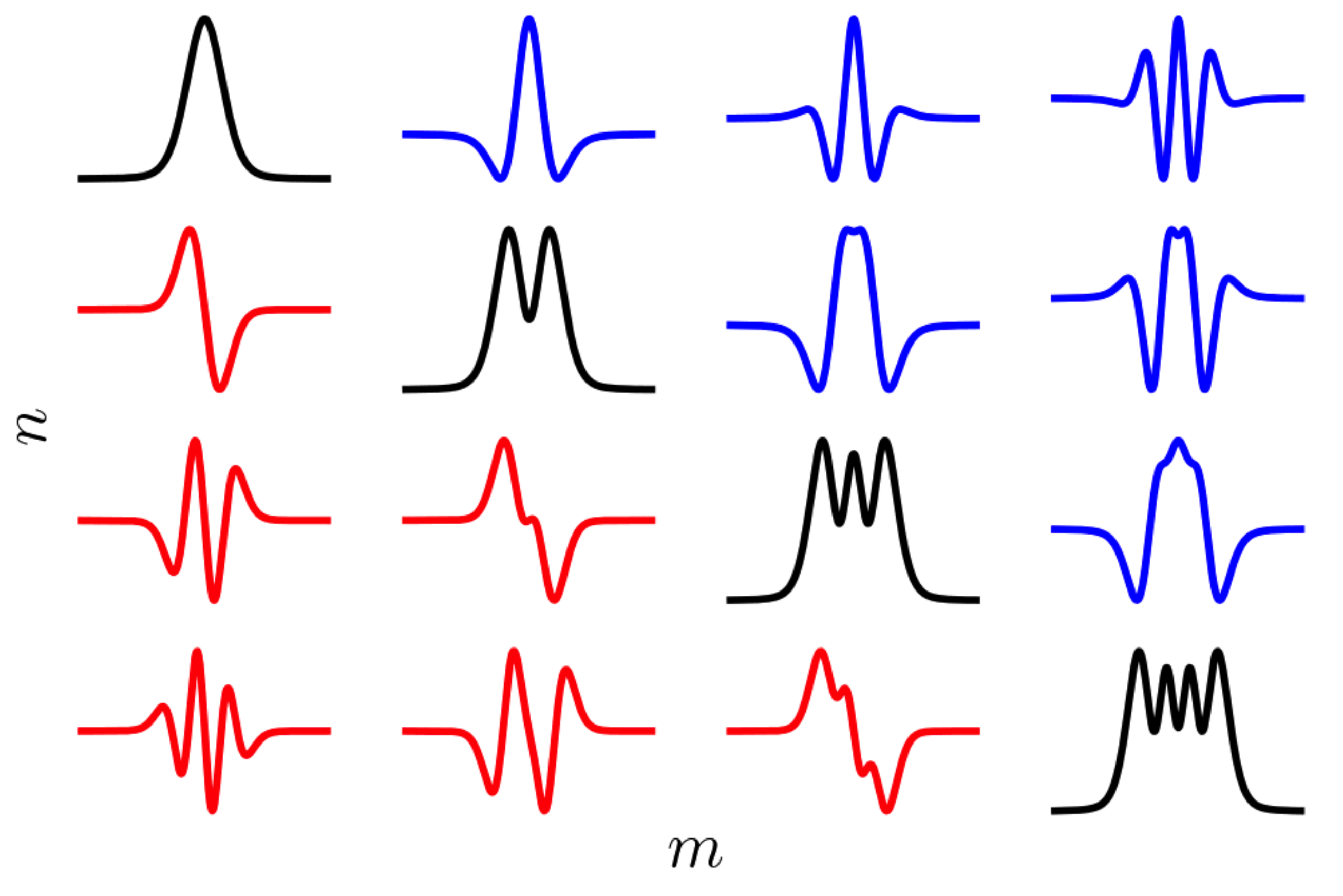}
\caption{\label{camel} Behavior of $\pi_{n,m}(p_z) $ for $1 \leq n,m \leq 4$. The diagonal cases $m=n$ are real functions shown as black curves. The non diagonal cases are complex functions, with their real and imaginary parts plotted respectively in blue above the diagonal, and in red below the diagonal. [Colors on line]} 
\end{figure}

The functions $\pi_{n,m}(p_z) $ are represented in figure \ref{camel} for the first values of $n$ and $m$. The diagonal cases $m=n$ are real-valued functions $\vert\widetilde{\psi_n}(p_z)\vert^2$ shown as black curves. The non diagonal cases $m\neq n$ give complex functions, their real parts are plotted as blue curves above the diagonal, while their imaginary parts are plotted as red curves below the diagonal (exchanging the roles of $n$ and $m$ corresponds to complex conjugation).

The diagonal cases show \emph{multi-bumps camel} shapes corresponding to autocorrelation functions of the Fourier transform of the Airy function ($n$ bumps for $\pi_{n,n} $). Meanwhile, the off-diagonal terms contribute as oscillations with time frequency $\omega_{nm} \equiv (\lambda_n-\lambda_m)/t_g$ in the expansion of $\Pi_t(p_z)$ with cross-correlation functions having more complex forms and non-null phases. The final detection picture discussed below reveals this complex interference pattern which depends on the value $g$, allowing one to estimate $g$ from the observed pattern.

\begin{figure}[ht]
\centering
\includegraphics[width=1\linewidth]{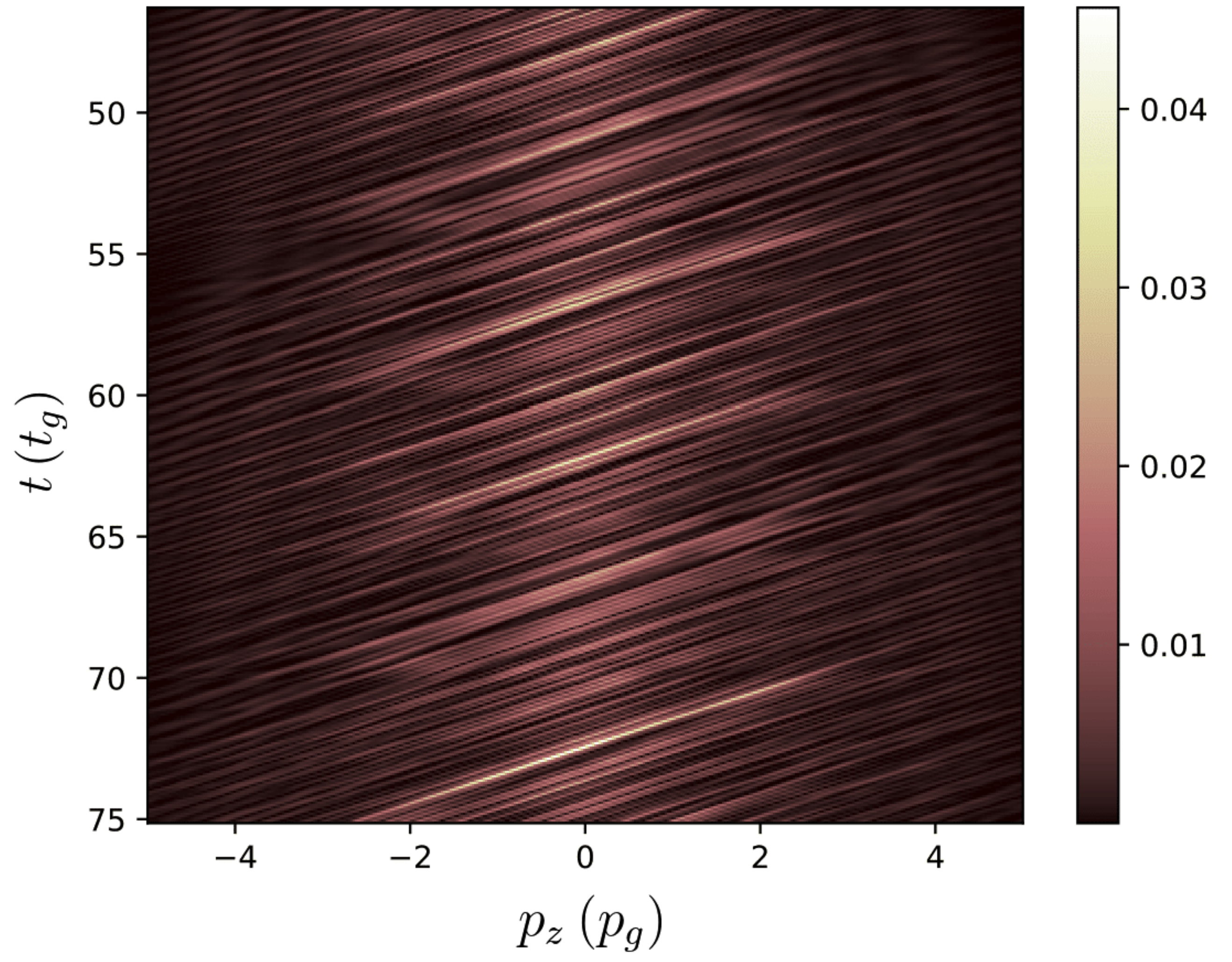}
\caption{\label{mirror} Probability density in momentum space $\Pi_t(p_z)$ at the end of the interference zone for $\zeta = 0.5\,\upmu$m, $h = 10\,\upmu$m and $v_0 = 0.8\,\text{m.s}^{-1}$ as a function of $p_z$ (horizontal axis, unit $p_g$) and $t$ (vertical axis, unit $t_g$; unit chosen for the distribution so that it is normed).  [Colors on line]}
\end{figure}

At this stage, it is worth showing the function \eqref{momentumpro} plotted in figure \ref{mirror} for the values of the initial parameters discussed above. It corresponds to interference of hundreds of different GQS, with a high weight for the lower-lying energy states.  The resulting probability density in momentum representation shows quasi-periodic oscillations with a somewhat complex shape. The abrupt transitions from negative $p_z$ to symmetrically positive $p_z$ correspond to bounces. The time bounds are chosen to delimit the portion of the signal that partakes in the observed interference pattern on the detection plate. 

We now compute the current probability density $J(X,T)$ on the horizontal detection plate where $X$ and $T$ are  the positions in space and time of the detection event.  We assume for simplicity that antiatoms are annihilated with 100\% probability at the detection plate as their kinetic energy is large after a free fall height $H=30$cm. The current $J(X,T)$ is written by using the Wigner function \cite{Wigner1932,Berry1977}. The free fall evolution of this function is classical as the potential varies linearly with $z$, which implies the following relations \cite{Dufour2015}~:
\begin{eqnarray}
&&\begin{split}
J(X,T)&=\int_{\mathbb{R}^2}\text{d}P_x\text{d}P_z\frac{P_z}{m}\,W_T(X,Z,P_x,P_z)  \\
&=\int_{\mathbb{R}^2}\text{d}p_x\text{d}P_z\frac{P_z}{m}\,W_t(x,z,p_x,p_z) ~,
\end{split}  \notag \\
&&\begin{split}
X &= x + \frac{p_x\tau}{m} \;,\quad
Z = z + \frac{p_z\tau}{m} - \frac{g\tau^2}{2} ~,  \\
P_x&= p_x \;,\quad P_z = p_z -mg\tau \;,\quad \tau\equiv T-t 
\end{split} 
\end{eqnarray}
with $W_T$ the Wigner function at time $T$.

These relations can be solved for parameters $t,x,z,p_x,p_z$ in terms of those 
associated with the detection event. A second degree equation 
has to be solved to extract the free fall time $\tau$, but
the extraction may be simplified by using the fact that 
the free  fall height $H$  is much  larger  than all typical length scales $h,\zeta$ at the end
of the interference zone and, consequently, the final momentum $P_z$ is much larger
than the typical momentum at the end of the interference zone.

\begin{figure}[ht]
\centering
\includegraphics[width=1\linewidth]{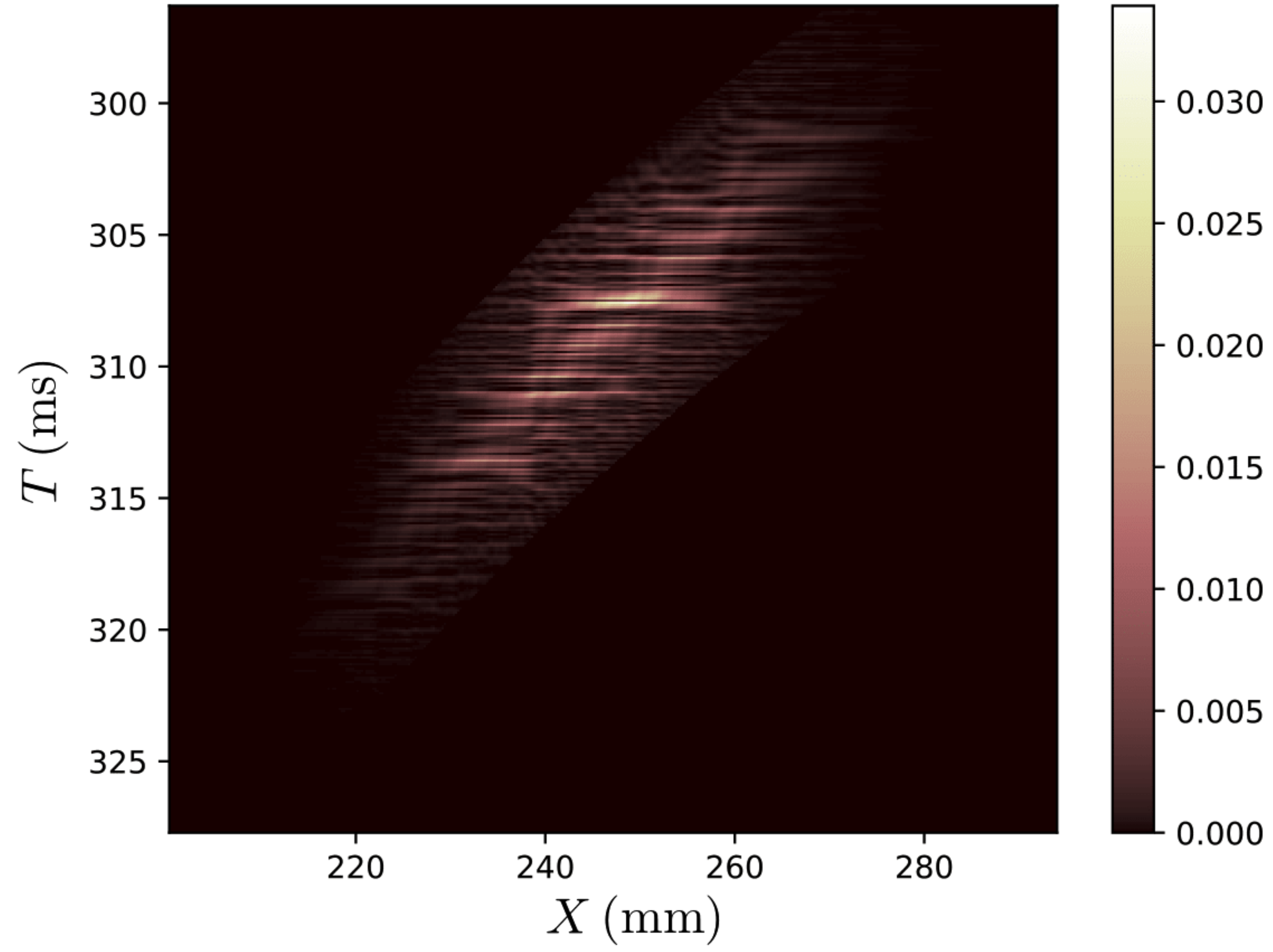}
\caption{\label{det} Probability current density  $\vert J(X,T)\vert$ on the detection plate for $\zeta = 0.5\,\upmu$m, $h = 10\,\upmu$m and $v_0 = 0.8\,\text{m.s}^{-1}$ as a function of $X$ (horizontal axis, in mm) and $T$ (vertical axis, in ms; unit chosen for the distribution so that it is normed).  [Colors on line]} 
\end{figure}

Using this macroscopic approximation for the free fall to the detection zone, we finally write $J(X,T)$ 
from the marginal of the Wigner function after the interference zone, integrated over space variable,
that is also the probability in momentum space $\Pi$~:
\begin{align}
&\vert J(X,T)\vert \equiv \frac{gm^2}{\overline{\tau}} \, \vert\widetilde{\phi_t}\left(p_x\right) \vert^2 \Pi_{t}(p_z) ~,
\notag \\
	&t = \overline{\tau} \frac{d}{X-d} \;,\quad \overline{\tau}\equiv \sqrt{\frac{2H}{g}} ~,  \\
	&p_x= \frac{m(X-d)}{\overline{\tau}} \;,\quad  p_z = mg\left(T - \frac{\overline{\tau} X}{X-d}\right) ~.\notag
\end{align}

The probability density $\vert J(X,T)\vert$ ($J(X,T)<0$ for freely falling atoms) is given 
from the probability density $\Pi_{t}(p_z) $ through 
a simple anamorphosis and a weighing by the probability density of the horizontal momentum 
$\vert\widetilde{\phi_t} \vert^2$.
The resulting  $\vert J(X,T)\vert$ is plotted in figure \ref{det} for the same parameters as in figure \ref{mirror}. 
The units are given in millimeters for $X$ and milliseconds for $T$ with these scales showing clearly that the event 
detection resolution, of the order of 0.1 mm in space and 0.1 $\upmu$s in time in the current GBAR design, 
is largely sufficient for getting the interference pattern in its full details. 

\begin{figure}[th]
\centering
\includegraphics[width=0.8\linewidth]{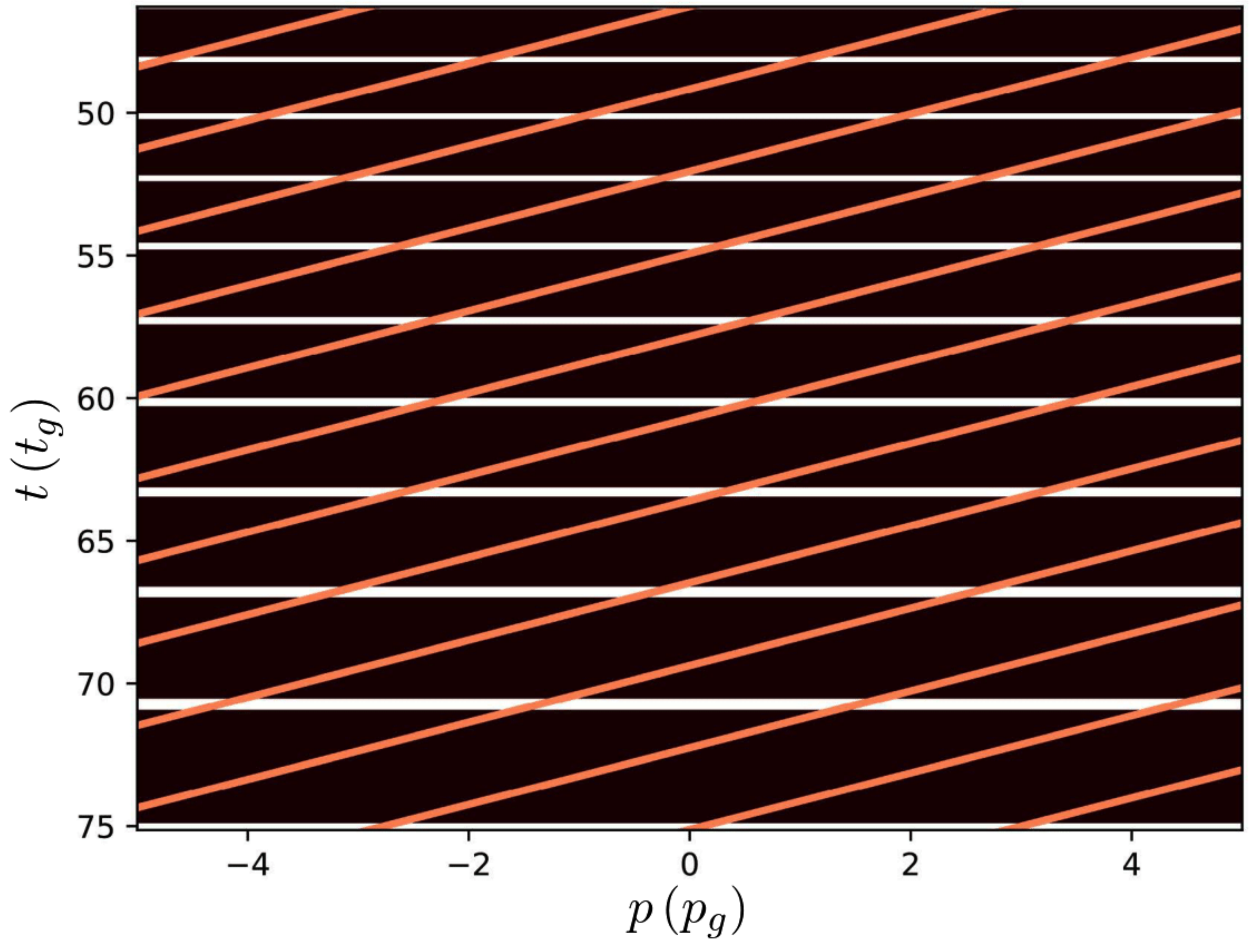}
\includegraphics[width=0.8\linewidth]{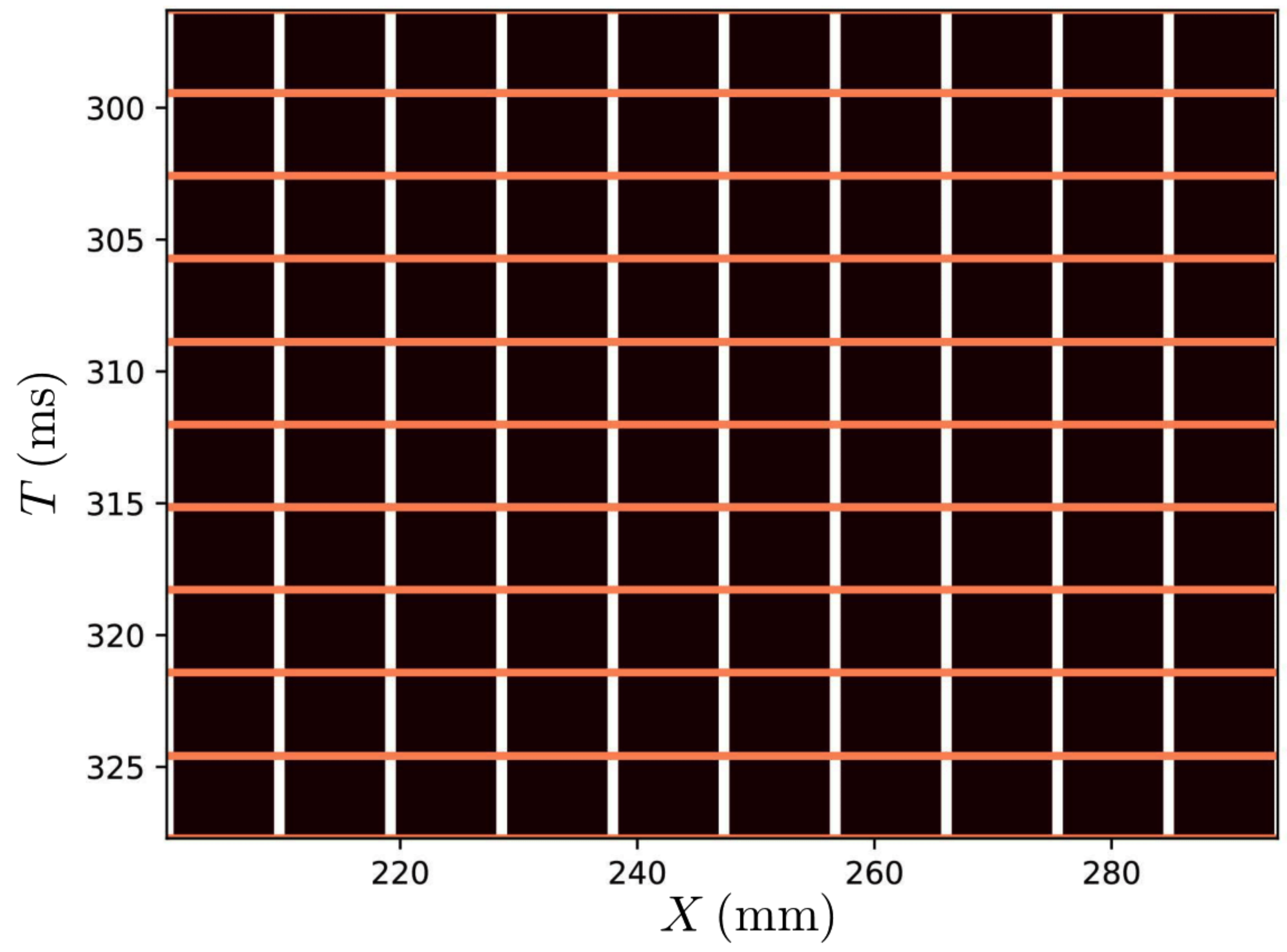}
\caption{\label{grids} Grids of lines transformed into one another by the anamorphosis at the endside of the mirror (upper plot) and on the detection plate (lower plot). The orange lines represent the constructive interferences, they correspond to classical free-fall trajectories on the mirror and to horizontal lines on the detector ($\delta T=0$). The white lines are horizontal on the upper plot  ($\delta t=0$) and vertical on the lower plot ($\delta X=0$). The bounds for the two plots are the same as on figure \ref{mirror} and figure \ref{det} respectively. [Colors on line]} 
\end{figure}

In order to understand the relationship between the two figures \ref{mirror} and \ref{det}, it is worth looking at the anamorphosis relations when fixing $T$ or $X$, and observing the resulting variations of $t$, $p_x$ or $p_z$ :
\begin{align}
\label{anamorphose}
&\delta X =0 \; \rightarrow \; \delta t =  \delta p_x = 0 \;,\quad \delta p_z = mg\delta T \quad;   \\
&\delta T =0 \; \rightarrow \; \frac{\delta t}t = \frac{\delta p_x}{p_x} = - \frac{\delta X}{X-d} 
\;,\quad \delta p_z = -mg\,\delta t \;. \notag 
\end{align}
The bright oblique lines corresponding to constructive interferences and classical free-fall movements in figure \ref{mirror} become the bright horizontal lines in figure \ref{det} which correspond to constructive interferences and are parallel to the $X$ axis on the detector plate. This discussion is illustrated by the two plots in figure \ref{grids} where
orange lines represent constructive interferences transformed into one another by the anamorphosis. Meanwhile, white lines, also transformed into one another by the anamorphosis, are vertical on the detector ($\delta X=0$) and horizontal on the plot corresponding to the end of the mirror ($\delta t=0$).

We also stress at this point that the positions in $X$ and $T$ on the detection pattern are directly measuring the momenta along
the two axis in the initial distribution. Analysing the envelop of the detection pattern thus allows one to assess the parameters
of this distribution, which would have to be done as the first step of the analysis for real experimental data.
Here we go on by considering that the initial distribution is known.

\paragraph{Uncertainty estimation}
We now estimate the uncertainty in the estimation of the value of $g$ from the interference pattern registered on the detection plate. 

As the distance between fringes depends on $g$, it could be tempting to measure $g$ directly from it. This technique is however unpractical here, because we have only a small number of annihilation events to sample the details of the probability distribution.  We use a much more robust maximum likelihood method to estimate the parameter $g$ and then deduce a variance for this estimation. 

We assume that we have 1000 prepared $\overline{H}$-atoms that is also $N=800$ detection events with $20\%$ $\overline{H}$-atoms lost in the absorber. We thus draw randomly 800 detection events in the probability distribution $\Prob_{g_0}$ corresponding to an \textit{a priori} value of the acceleration, say the standard value $g_0$. 
We consider that this random draw of detection events $\Draw=\left\{(X_i,T_i),\;1 \leq i \leq N\right\}$ simulates the output of one experiment. 

We then use a maximum likelihood method to get an estimator $\hat{g}$ of the parameter $g$ as would be done in the data analysis of the experiment. This estimator $\hat{g}$ maximizes the likelihood of the random draw  $\Draw$ to reproduce the distributions $\Prob_{g}$ corresponding to different \emph{a posteriori} values of the parameters~: 

\begin{align}
\Like_\Draw\left(g\right) &= \prod_{i=1}^N \Prob_g(X_i,T_i) \; , \quad \ln\Like_\Draw\left(g\right)=\sum_{i=1}^N \ln\Prob_g(X_i,T_i)  \notag \\
 &\left( \frac{\partial \ln\Like_\Draw\left(g\right)}{\partial g}  \right)_{\hat{g}}=0~.
\end{align}

\begin{figure}[ht]
\centering
\includegraphics[width=1\linewidth]{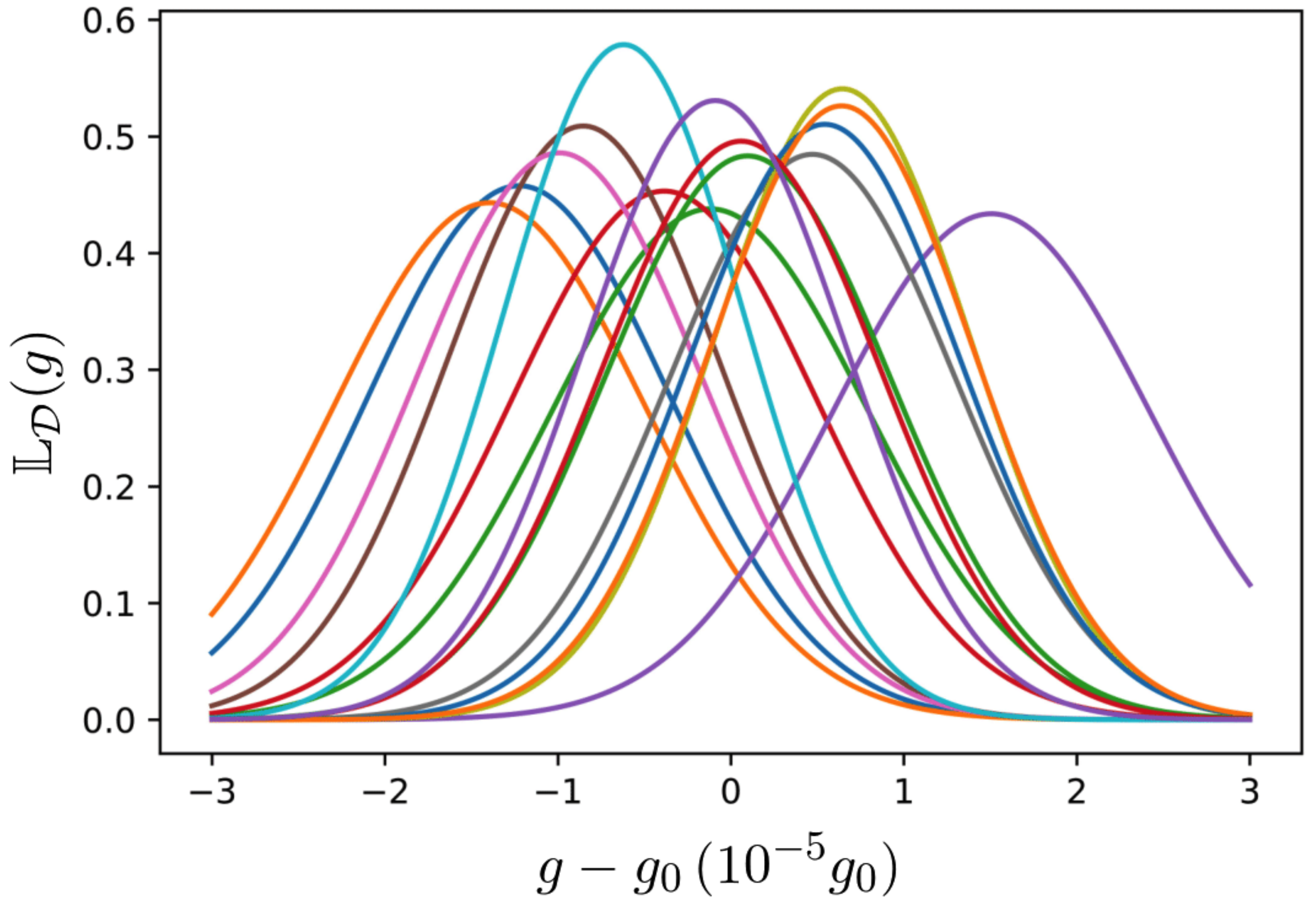}
\caption{Gaussian distributions obtained by a quadratic fit of the log-likelihood function calculated for 15 different random draws of $N=800$ atoms. The gaussians are normalized so that the variation of their variance is seen more easily as a variation of their height. The colors have no meaning, they only allow one to distinguish the various functions.The horizontal axis scales as $(g-g_0)/g_0 \times 10^5$.  [Colors on line]}
\label{gauss}
\end{figure}

Figure \ref{gauss} shows quadratic fits of the log likelihood functions (i.e. gaussian fits of the likelihood functions) around their extrema. These fits correspond to 15 random draws of 800 events, with each fit yielding an estimator $\hat{g}$ of the parameter $g$ and an estimator  $\hat{\sigma}_g$ of the dispersion associated with this estimator :
\begin{align}
& \ln \,\Like_\Draw\left(g\right) \approx{} -a_\Draw g^2 + b_\Draw g +c_\Draw \;, \notag \\
&\hat{g} = \frac{b_\Draw}{2a_\Draw} \;,\quad \hat{\sigma}_g = \frac1{\sqrt{2a_\Draw}} \;.
\end{align}

We have normalized the gaussians so that the variation of their variance is seen more easily as a variation of their height. The variation of the peaks shows the dispersion of the estimator $\hat{g}$ around $g_0$ for different random draws. The number of 15 draws has been chosen to illustrate this variance while simultaneously avoiding confusion on the figure. 
The variation of $\hat{g}$ and the value $\hat{\sigma}_g$ are estimators of the uncertainty in the measurement of $g$, which both tend to be reliable if the statistical efficiency of the method is good \cite{Cramer1946}.

The statistical efficiency of the method is indeed found to be quite good as the dispersions $\hat{\sigma}_g$ in figure \ref{gauss}  are close to the Cramer-Rao lower bound given by the Fisher information $\mathcal{I}_g$  in the detection pattern \cite{Cramer1946,Refregier2004} ($\EX$ used to denote expectation values, 
$\Sigma_g^2$ is the variance of $\hat{g}$) :
\begin{align}
\label{Fisher}
&\Sigma_g^2 \geq \frac1{N \Fish(g)}  \;,  \\
&\Fish(g) =  \EX\left[- \frac{\partial^2}{\partial g^2} \ln\Prob_g \right]
=  \EX\left[ \left(\frac{\partial}{\partial g} \ln\Prob_g\right)^2 \right]  \;. \notag
\end{align}
The second expression of the Fisher information $\Fish(g)$ given in \eqref{Fisher} explains why the precision in the estimation of $g$ is greater for interference patterns exhibiting fine details depending on the parameter $g$. In simple words, these details act as thin graduations that make it easier to observe small displacement and distortion of the interference pattern when $g$ is varied.

\begin{figure}[hb]
\centering
\includegraphics[width=1.0\linewidth]{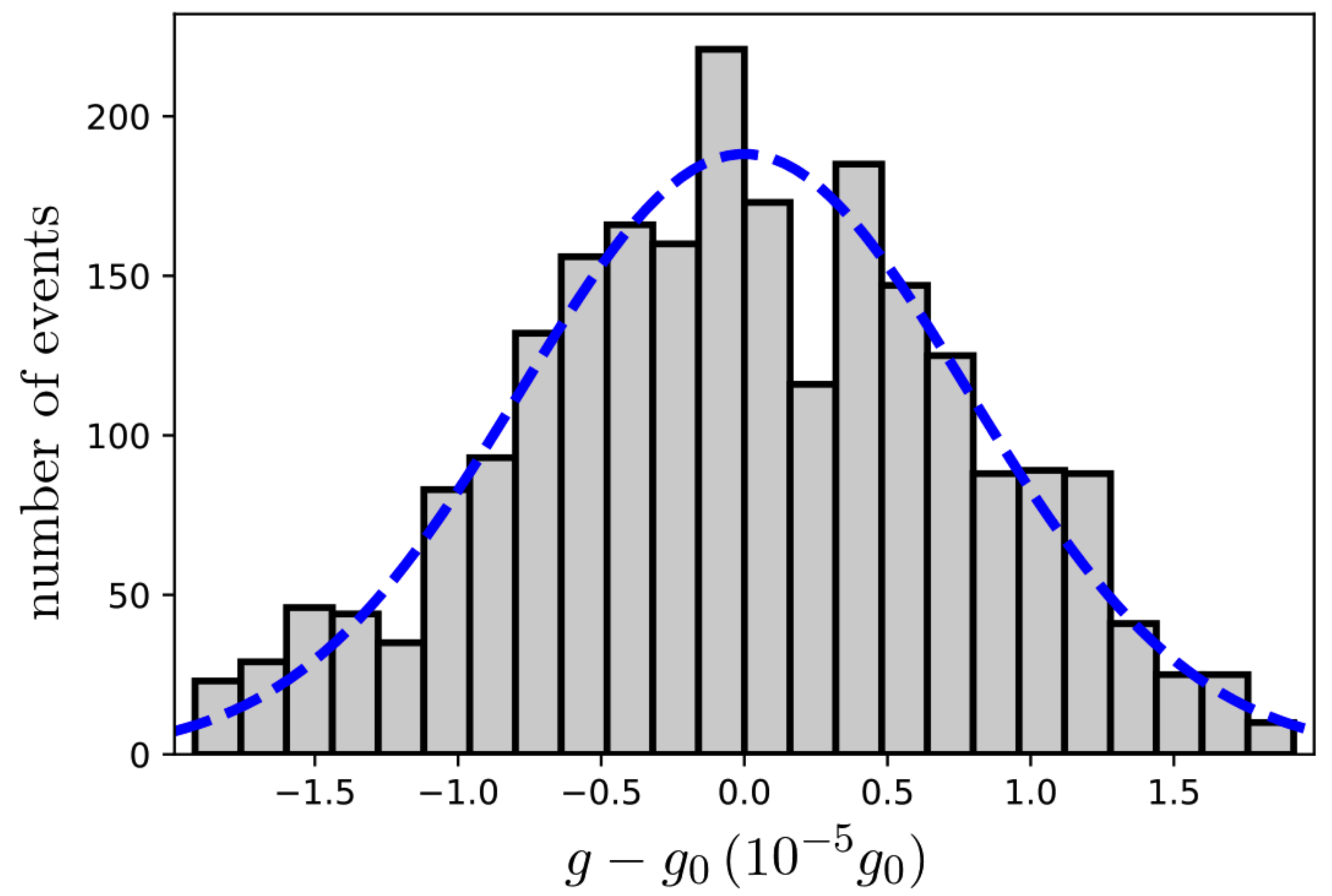}
\caption{\label{histogram} Histogram of the relative variations $(\hat{g}-g_0)/g_0 \times 10^5$ obtained by repeating 2300 times a Monte Carlo simulation on 800 events, for $\zeta = 0.5~\mu$m, $h = 10~\mu$m and $v_0 = 0.25~\text{m.s}^{-1}$. The vertical axis counts the number of events per channel. The blue dashed line is a gaussian fit of the histogram. [Colors on line] } 
\end{figure}

In order to give a robust estimation of the variance, we have finally repeated the full procedure for $M$ different random draws of the $N$ points. 
The histogram shown in figure \ref{histogram} corresponds to $M=2300$ such draws of the $N$ points, with the parameters $\zeta$ and $h$ corresponding to figure \ref{det}, and a velocity $v_0 = 0.25\,\text{m.s}^{-1}$.
The blue dotted line is a gaussian fit of the histogram which gives the dispersion $\Sigma_g$ now calculated on a large number $M$ of experiments repeated in the same conditions :
\begin{align}
\Sigma_g & \simeq 7.8 \cdot 10^{-6} \, g   \label{sigmag}
\end{align}

The dispersion $\Sigma_g$ is obtained by repeating a numerical experiment in conditions where the real experiment can not be repeated due to the small number of available $\overline{H}-$atoms.
It is more reliable than the value obtained directly on a single draw corresponding to a single experiment. Thanks to the good efficiency however, the expectation value of the estimator $\hat{\sigma}_g$ obtained in a single draw is close to it ($\EX(\hat{\sigma}_g) \simeq 7.5 \cdot 10^{-6} \, g \simeq \Sigma_g$) while its relative dispersion is small (dispersion of $\hat{\sigma}_g/\EX(\hat{\sigma}_g) \simeq 8\%$).

\paragraph{Discussions}
The calculations presented in this paper show that a large improvement of the $g-$measurement accuracy may in principle be attained by using quantum interference methods rather than classical timing.

We now compare quantitatively the uncertainty of the quantum interference method with that of the classical timing measurement corresponding to the current design of GBAR, by using the same parameters in the quantum and classical methods. To this aim, we repeat the same discussion as in the previous section for the classical experiment, 
which leads to a relative uncertainty on that measurement $\Sigma_g/g=1.7 \cdot10^{-3}$. 

The initial position dispersion $\zeta=0.07~\mu$m considered in the current  design of GBAR corresponds to a larger vertical velocity dispersion that dominates the relative uncertainty on that measurement and it leads to the relative uncertainty $\Sigma_g/g = 1.2\cdot 10^{-2}$ expected for this current design. 

The spectacular improvement of $\Sigma_g/g$ is due partly to the change of the initial dispersion but
the more important effect is associated with the change from a classical measurement to a quantum one. The interpretation of this change is intuitively clear as the quantum interference pattern contains much more information than the classical one, which explains why the uncertainty in the estimation of $g$ is much better. 

As discussed above, this analysis is reliable as long as the variance obtained in the Monte-Carlo simulation is close to the Cramer-Rao lower bound. When the number $N$  of atoms decreases, the distance of the two values increases, meaning that the statistical efficiency is degraded. Our simulations show that the efficiency would remain good enough if the number of available atoms was smaller than currently expected, so that the precision of the quantum method would remain by far better than that of the classical method.

A smaller kick velocity would enhance the duration of the interference period above the mirror while also increasing  the probability for the atom to be annihilated. For a realistic treatment of the uncertainty calculation, we should thus describe the reflection on the surface by adding an energy dependent annihilation probability at each bounce of the atom above the surface \cite{Crepin2017}.
It would also be necessary to take into account quantum reflection on the detection plate \cite{Dufour2013}. 

Of course, other improvements would be required to build up a reliable data analysis of the experiment when needed, as many details have been omitted in the preliminary analysis presented in this paper. We feel that they would not change the main conclusion of this paper, namely that the quantum interference technique  opens attractive perspectives for more accurate equivalence principle tests on antihydrogen atoms.

\paragraph{Aknowledgments}
Thanks are due to P. Clad\'e and M.-P.  Gorza for insightful discussions about uncertainty estimation
and also to our colleagues in the GBAR and GRANIT collaborations.

%%% Bibliography
%\bibliography{bibliography}
%\bibliographystyle{unsrt}

\end{document}